**Photoacoustic Silk Scaffolds for Neural stimulation and Regeneration**

*Nan Zheng, Vincent Fitzpatrick, Ran Cheng, Linli Shi, David L. Kaplan, and Chen Yang[*]*

N Zheng, Prof. C Yang

Division of Materials Science & Engineering, Boston University, Boston, MA 02215, USA

Dr. V. Fitzpatrick, Prof. D. L. Kaplan

Department of Biomedical Engineering, Tufts University, Medford, MA 02215, USA

R. Cheng, L. Shi, Prof. C. Yang

Department of Chemistry, Boston University, Boston, MA 02215, USA

Prof. C. Yang

Department of Electrical and Computer Engineering, Boston University, Boston, MA 02215, USA

*corresponding author: E-mail: cheyang@bu.edu

**Abstract**

Neural interfaces using biocompatible scaffolds provide crucial properties for the functional repair of nerve injuries and neurodegenerative diseases, including cell adhesion, structural support, and mass transport. Neural stimulation has also been found to be effective in promoting neural regeneration. This work provides a new strategy to integrate photoacoustic (PA) neural stimulation into hydrogel scaffolds using a nanocomposite hydrogel approach. Specifically, polyethylene glycol (PEG)-functionalized carbon nanotubes (CNT), highly efficient photoacoustic agents, are embedded into silk fibroin to form biocompatible and soft photoacoustic materials. We show that these photoacoustic functional scaffolds enable non-genetic activation of neurons with a spatial precision defined by the area of light illumination, promoting neuron regeneration. These CNT/silk scaffolds offered reliable and repeatable photoacoustic neural stimulation. 94% of photoacoustic stimulated neurons exhibit a fluorescence change larger than 10% in calcium imaging in the light illuminated area. The on-demand photoacoustic stimulation increased neurite outgrowth by 1.74-fold in a dorsal root ganglion model, when compared to the unstimulated group. We also confirmed that photoacoustic neural stimulation promoted neurite outgrowth by impacting the brain-derived neurotrophic factor (BDNF) pathway. As a multifunctional neural scaffold, CNT/silk scaffolds demonstrated non-genetic PA neural stimulation functions and promoted neurite outgrowth, providing a new method for non-pharmacological neural regeneration.

1. **Introduction**

In tissue engineering, scaffolds offer multiple synergistic functions for regeneration, including providing support for cell adhesion, structural support for the tissue, and facilitating mass transport of nutrients and growth factors, while also being biocompatible. For neural tissue, biophysical and biochemical stimulation leveraging the innate ability of neural tissue to react to stimuli is an additional key strategy to improve neurite outgrowth and promote functional regeneration[1]. Approaches to functionalized nerve scaffolds include physical stimulation, through electrically conductive scaffolds,[2] mechanical cues[3], and chemical stimulation, through drug and growth factor release[4].

By modulating neuronal activities, neural stimulation techniques like optogenetics[5], electrical[6] and ultrasound stimulation[7], encourage nerve regeneration. Electrical stimulation is the most widely applied neural stimulation technique to promote nerve regeneration. For example, applying an external electric field for 1 hour at 20 Hz in vitro significantly promoted axon regeneration after nerve transection and microsurgical repair[8]. Conventional ultrasound treatment can be non-invasive but has limited spatial resolution. Optogenetics promoted neurite outgrowth with high spatial precision and in a cell-specific manner[5]. However, this approach requires viral transfection which makes human applications challenging.

Electrical stimulation has also been integrated with various conductive scaffolds. These scaffolds showed remarkable performance in promoting nerve regeneration, including conductive polymers[9] and carbon-based materials[2a, 10]. However, in clinical applications, the delivery of electrical stimulation to conductive scaffolds remains challenging. Current solutions are limited to stimulation within the intraoperative time frame, or transcutaneous stimulation after surgery. Multiple postoperative stimulations are important for enhanced neuron regeneration and functional recovery[11], but require the use of transcutaneous wires, and therefore an increased risk of infection. Also, the electrical current applied for stimulation can spread beyond the injury site to healthy tissues, with undesirable impact.

Recently, photoacoustic (PA) stimulation has been shown as a non-genetic and high-precision method for neural stimulation[12]. In a photoacoustic process, pulsed light is delivered into absorbers, generating acoustic waves at ultrasonic frequencies [13]. Photoacoustic polymer nanoparticles, upon excitation by pulsed light in the second near-IR window, activate primary neurons in culture with single-neuron precision[12b]. Optical fibers with a photoacoustic coating composed of graphite mixed with epoxy, or carbon nanotubes mixed with PDMS, provide sub-millimeter photoacoustic stimulation in vitro and in vivo[14]. We hypothesized that introducing highly efficient absorbers like carbon nanotubes into a biocompatible material, through a nanocomposite approach, would enable PA stimulation in biomaterials used for scaffolds. As a light-mediated non-genetic method, neural scaffolds with PA stimulation capabilities could promote neurite outgrowth as an alternative to electrical stimulation and optogenetics, without the requirement of transcutaneous wire connections or gene therapy.

In this work, we report a biocompatible silk fibroin (silk) scaffold with PA capabilities, enabling neural stimulation and promoting neurite outgrowth. We introduced CNTs into silk using a nanocomposite approach. The silk matrix in the CNT/Silk material provides structural support for tissue growth. The embedded CNTs absorb NIR-II pulsed light and transduce light energy to acoustic, which activates neurons cultured on the CNT/Silk to promote neural outgrowth. We chose silk fibroin as a matrix material, as silk is an FDA-approved biocompatible material, and its abilities in supporting neural adhesion and preserving neural functions have been confirmed previously [15]. In addition, silk fibroin provides a controllable rate of biodegradation[16], tunable drug-loading capabilities[17], and tunable mechanical properties[18], all of which are important for neural scaffolds. CNTs provide high PA-conversion efficiency and strong absorption in NIR-II [19], which provides potential for tissue penetration.

Unlike our previous report on fiber-based PA devices, the generalizable scaffold approach described here in film or 3D structure formats, offers new implant materials for neural stimulation and enhanced regeneration. PA stimulation achieved upon the excitation of the NIR light opens up potential non-invasive neural stimulation and thus regeneration. Our approach suggests a new neural interface based on PA biomaterials and highlights the efficacy of PA stimulation for regeneration.

**Results and Discussion**

**2.1 Fabrication and characterization of CNT/Silk scaffold**

The CNT/Silk scaffolds are nanocomposite materials consisting of silk fibroin and CNTs. We functionalized the CNTs with DPSE-PEG2000 (1,2-distearoyl-sn-glycero-3-phosphoethanolamine-N-[methoxy(polyethylene glycol)-2000]) to uniformly disperse the CNT in the silk and to increase the concentration of CNTs in the silk, in order to maximize the light-to-acoustic conversion [20] **(Figure S1)**. To prepare the CNT/silk scaffolds, 20-100 µg/mL of functionalized CNT solution was mixed with 2% (w/v) silk fibroin solution to obtain the solution for casting. The CNT/Silk scaffolds were then prepared through a cast-and-dry method **(Figure 1a)** to form films with a typical thickness of 32.54 ± 1.98 µm **(Figure. S2a)** or a 3D roll structure **(Figure S2b)** through the self-folding strategy we have previously reported[21]. All CNT/silk films used in following experiments were fabricated from solutions of 20 ug/mL functionalized CNT and 2% (w/v) silk fibroin, unless otherwise specified.

We quantified the PA signals generated by the CNT/Silk films under pulsed laser illumination. A 1030 nm laser with a pulse width of 3 ns at a 1.7 kHz repetition rate was delivered through a 200 µm core diameter multimodal optical fiber, providing an illumination area of 0.05 mm$^2$ to the CNT/Silk film. The acoustic waveforms generated were measured by a 10 MHz ultrasound transducer under water. The schematic of the measurement is shown as **Figure 1b**. A representative acoustic wave in the time domain under a single laser pulse is shown in **Figure. 2c**. After Fast Fourier Transform (FFT), the PA frequency spectrum exhibited a broad frequency range up to 20 MHz and a peak frequency of 1.2 MHz. This broad frequency band is characteristic of PA signals, compared to narrow frequency bands in ultrasound generated by traditional piezoelectric-based transducers[22].

To control the amplitude of the PA signals generated, we varied the concentration of CNTs in the composite, as well as the laser energy, based on the following rationale. According to the basic PA theory[23], the pressure (Pa) is correlated to several factors and can be expressed as:

$$P_0 = \frac{\beta c^2}{C_p} \cdot A \cdot \frac{F}{l} \qquad (\text{eq. 1})$$

where $\beta$ is the volumetric thermal-expansion coefficient (K$^{-1}$), $c$ is the sound speed (m s$^{-1}$), $C_p$ is the specific heat capacity at constant pressure (J kg$^{-1}$ K$^{-1}$), $A$ is the light absorption (0<A<1, dimensionless), $F$ is the laser fluence (J m$^{-2}$) and $l$ is the characteristic length (m). Therefore, the amplitude of the PA wave is linearly proportional to the light absorption and input laser energy. We first varied the light absorption of the PA film by changing the concentration of the embedded CNTs **(Figure 1d)**. The CNT/Silk films fabricated from 0, 20, 40 and 80 µg/mL functionalized CNTs and 2% (w/v) silk fibroin were tested with a fixed laser pulse energy of 74.7 µJ. Second, we tested the effect of laser pulse energy. The CNT/Silk film with a CNT concentration of 20 µg/mL was used, and the laser pulse energy varied from 22.9 to 74.7 µJ. In both experiments, the higher concentration of embedded CNTs and larger laser pulse energy induced larger amplitudes of PA waves, in a linear fashion (R-Square = 0.978 and 0.972, respectively), as expected based on Equation (1). The linear characteristics enabled facile control of the amplitude of PA waves through these two key parameters (CNT concentration and laser pulse energy). It should be noted that a 40 µm needle hydrophone was also used to measure the pressure of the PA waves from a 20 µg/mL CNT/Silk film at a pulse energy of 14.7 µJ (corresponding to an energy density of 29.4 mJ/cm$^2$). The generated PA pressure was 0.185 ± 0.015 MPa, which will be shown below to successfully induce neural stimulation in primary neuron cultures.

The photoacoustic effect is associated with a transient temperature increase. To characterize the thermal increase generated by CNT/Silk scaffolds during the photoacoustic process, the temperature of the CNT/Silk surface of the light illumination area was measured by a thermal camera. The laser pulse energy and repetition rate were fixed at 53 µJ and 1.7 kHz. The laser duration was increased from 3 ms to 100 ms. **Figure 1f** plots the temperature measured as a function of time upon the excitation of a laser pulse train with a duration at 3, 10, 20, 50 and 100 ms, respectively. A maximal temperature increase of 0.64 ± 0.06 K was observed for a laser duration of 100 ms. For durations of 50 ms and 20 ms, the temperature increases were 0.42 ± 0.15 K and 0.26 ± 0.03 K, respectively. A negligible temperature increase was observed when the duration was below 10 ms for the CNT/Silk scaffold. No temperature increase was noticed on the surface of the pure silk scaffold, confirming that the observed temperature increase was mainly due to the light absorption of CNTs. These results suggest that a minimal thermal effect was generated during the photoacoustic process with the conditions tested. Specifically, when the photoacoustic process was performed with a laser pulse energy of 53 µJ and a laser duration below 10 ms, no photothermal effect was exhibited, suggesting that cells will not be negatively impacted due to thermal injury during PA stimulation of the scaffolds.

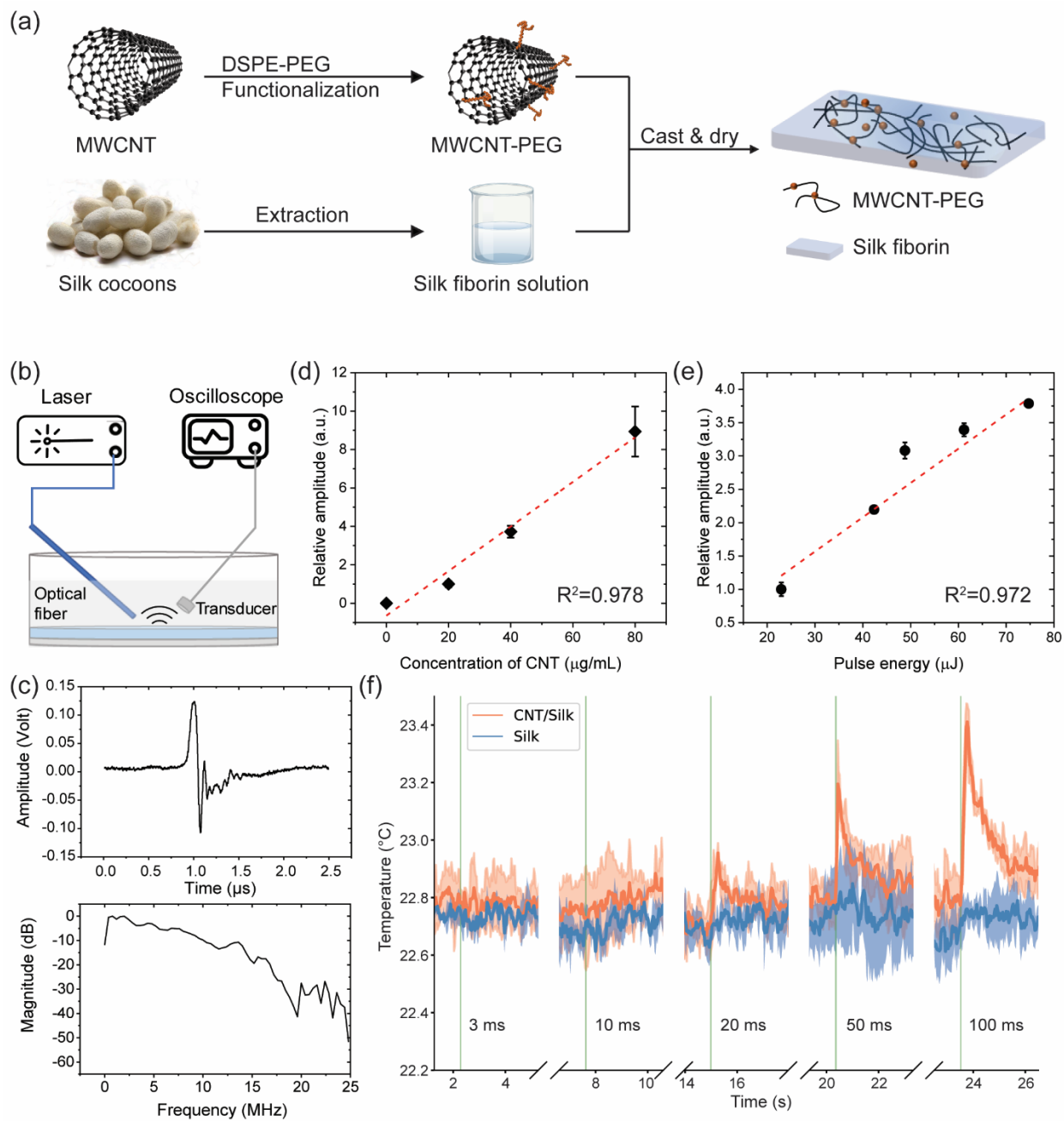

**Figure 1. Fabrication and characterization of CNT/silk PA films.** (a) Schematic of the fabrication process. (b) Schematic of the PA measurement. (c) Acoustic wave generated by a 20 µg/mL CNT/silk film in time domain and frequency domain measured by a 10 MHz transducer. Laser pulse energy of 74.7 µJ. (d, e) Normalized mean peak to peak amplitude of the PA waves as a function of the concentration of CNT (d) and laser pulse energy (e). Error bar, n = 3. Red dotted lines, fitting curves. (f) Temperature as a function of time measured at the surface of the area under laser illumination for silk film and CNT/Silk film. Laser durations of 3 ms, 10 ms, 20 ms, 50 ms and 100 ms. Laser pulse energy 53 µJ. Green solid lines indicate laser onset. n = 3, shaded area: ± standard error of the mean (s.e.m.).

## 2.2 In vitro and in vivo assessment of CNT/Silk scaffold biocompatibility

We tested the biocompatibility of the CNT/Silk PA materials both in vitro and in vivo. We first compared the morphology of rat cortical neurons cultured on CNT/silk films and glass (control group) through immunofluorescent staining and confocal microscopy **(Figure 2a)**. After 7 days of in vitro (DIV) culture, neurons were fixed and labeled with anti-Tau antibody (green) and DAPI for cell nuclei (cyan) to estimate total neurite length and quantify neuron population from an area of 0.1013 mm$^2$ for all samples **(Figure S3).** According to the literature and our previous experience, seeding density affects neuron growth [24]. Therefore, cultures with two seeding densities were studied: $1\times10^5$/cm$^2$, termed "low density", and $2\times10^5$/cm$^2$, "high density". Compared to the control group, neurons cultured on the CNT/silk films showed no significant differences in neurite length and number of neurons in both the high and low seeding density conditions **(Figure 2b, c)**. The viability of cortical neurons cultured on the CNT/Silk material was also evaluated using the MTS (3-(4,5-dimethylthiazol-2-yl)-5-(3-carboxymethoxyphenyl)-2-(4-sulfophenyl)-2H-tetrazolium) assay. In addition to the CNT/silk films and the unloaded silk films controls, we also evaluated unloaded silk films with increasing concentrations of freeform functionalized CNTs in the medium, to mimic degraded CNT/silk scaffolds. The concentration of CNTs was calculated from a previously reported degradation rate of 20% remaining weight after 14 days of enzymatic degradation from the silk film prepared by the same protocol[25]. After DIV14, the viability of neurons in each group was determined by MTS assay. No significant difference was detected in the CNT/silk film group, or the unloaded silk film with increased amount of CNT group, compared to the control group **(Figure 2c)**. These results confirmed that the presence of the CNT/Silk scaffold did not change the morphology or the viability of cultured neurons, and demonstrated the cytocompatibility of CNT/silk scaffolds in neuron culture.

The biocompatibility of implanted CNT/silk scaffold was evaluated using a mouse model. CNT/silk rolls with a length of 7 mm, diameter of 2 mm and CNT concentrations of 50 µg/mL and 100 µg/mL were prepared and implanted subcutaneously. The acute and chronic inflammatory responses were studied at days 3 and 30 post implantation, respectively, by H&E staining of histological slices of the explanted scaffolds. In the day 3 samples **(Figure 2e-h)**, there was a focal disruption of the skin surface and an insertion tract through the skin layers that led to the implant. This insertion tract was composed of clear space and inflammatory cells. There was severe dermal hyperplasia adjacent to or overlying the implant, and there were multiple serocellular crusts on the skin surface. Some mild inflammation (mostly macrophages and neutrophils and small numbers of lymphocytes and multinucleate cells) was observed in the periphery of the CNT/silk roll. At day 30, the insertion tract was not evident, and the skin surface was intact. The epidermis was normal to mildly hyperplastic. At day 30 **(Figure 2i-l)**, there was a thick layer of fibroplasia and/or granulation tissue (fibroplasia and neovascularization) encapsulating the CNT/silk rolls in all sections. Only scattered inflammatory cells (mostly lymphocytes with fewer macrophages and multinucleated giant cells) remained. A small amount of cellular debris was observed between layers within the CNT/silk rolls. At 3 and 30 days, there was no difference in the inflammatory response between the 50 µg/ml and 100 µg/ml concentration of CNT groups. The inflammatory response in the acute (day 3) samples was consistent with what would be expected following surgical insertion of an implant and subsequent surgical closure and was cleared by 30 days. By 30 days, fibroplasia and granulation tissue surrounded the implant with very slight inflammation and minimal overall response, indicative of biocompatibility.

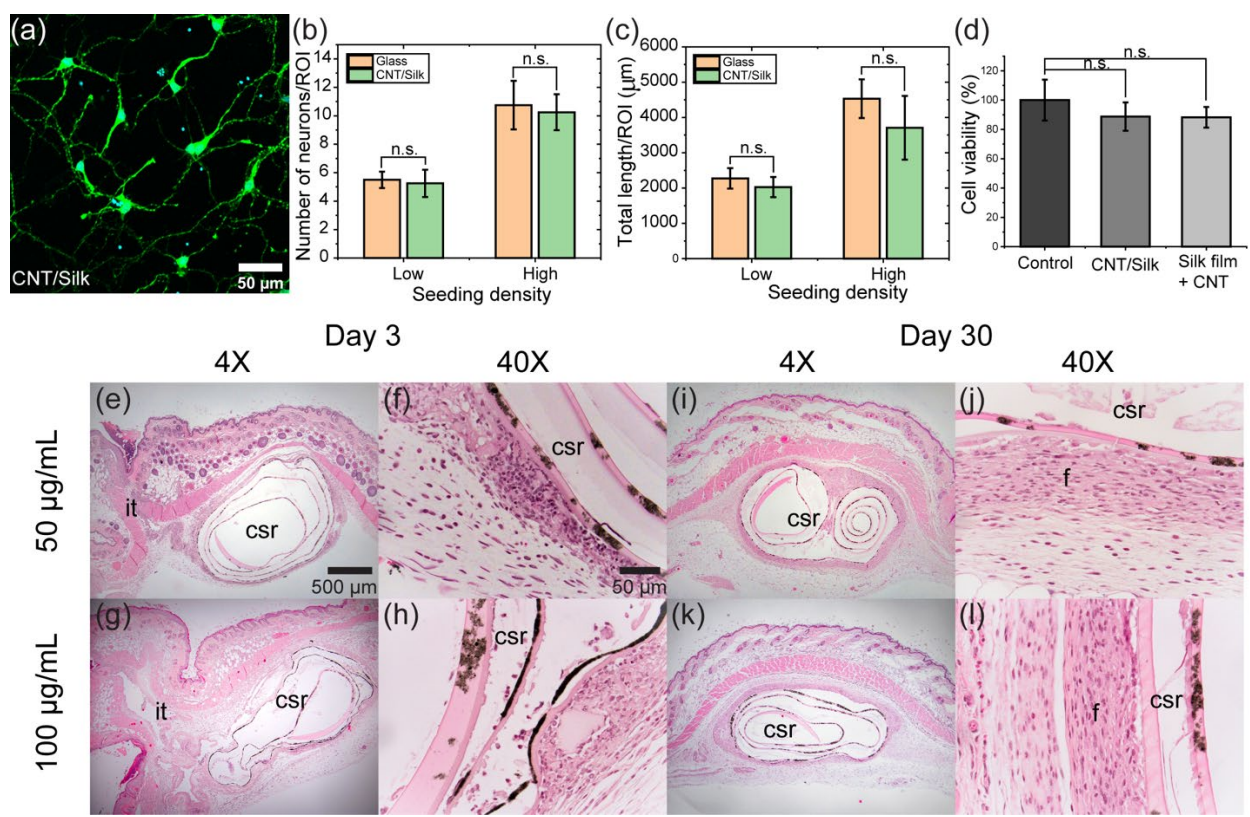

**Figure 2. Biocompatibility of CNT/Silk film evaluated in vitro and in vivo.** (a) Representative confocal image of cortical neurons cultured on CNT/Silk film. Neurons were fixed and stained with anti-Tau antibody (green) and DAPI for cell nuclei (cyan) at DIV7. (b-c) Quantification of the number of neurons per area (b) and the total neurite length per area (c). ROI: region of interest, 0.1013 mm$^2$ for all samples. The number of neurons was analyzed by counting the DAPI-stained nuclei and the total neurite length was analyzed by counting the length of anti-Tau labeled neurites. (d) Cell viability of neurons cultured on silk film as control, CNT/silk film, silk film with freeform CNT measured by MTS assay. Error bars represent standard deviation (n = 5, n.s., $p = 0.138$ (CNT/Silk), $p = 0.068$ (silk film + CNT), one-way ANOVA and Tukey's means comparison test). (e-l) Representative images of mouse skin with CNT/silk roll implants with different CNT concentrations. (e)-(h), Day 3. Mild inflammation; (i)-(l), Day 30. it: insertion tract; csr: CNT/silk roll; f: fibroplasia and granulation tissue.

## 2.3 CNT/silk scaffold stimulated neuronal activities through PA waves

To confirm the stimulation function of the CNT/Silk scaffolds, GCaMP6f (DIV 12-14) labeled primary neurons were cultured on the CNT/Silk films and calcium imaging was used to monitor neuronal activity **(Figure. S4).** A 3-nanosecond pulsed laser at 1030 nm with a repetition rate of 1.7 kHz was used to generate a laser pulse train with a duration of 5 ms, corresponding to 8 pulses with a pulse energy of 14.7 µJ (energy density = 29.4 mJ/cm$^2$) for PA neural stimulation. The laser light was delivered to the CNT/Silk surface though a 200 µm core diameter multimodal optical fiber. The location of the illumination area was controlled by a 3D micromanipulator, and the illumination area was calculated to be 0.05 mm$^2$. Representative fluorescence images of the neuron cultures on the CNT/silk films before and after PA stimulation are shown in **Figure 3a, b**, with the dashed circles showing the illumination area. Increased fluorescence intensity of GCaMP6f in individual neurons within the illumination area was clearly observed immediately after applying the pulsed laser. Of the 18 neurons studied, 17, corresponding to 94.4%, showed an increase in fluorescence greater $\Delta F/F_0$ than 10% after the onset of laser exposure (**Figure 3c,g**). $F_0$ is the baseline fluorescence signal of the neurons before the stimulation. The map of maximum fluorescence change $\Delta F/F_0$ (**Figure 3c**) also confirmed that neurons within the illumination area were successfully activated, while no fluorescence change was observed outside of the illumination area.

To confirm that the observed activation was due to PA stimulation rather than light illumination only, the calcium traces of PA-stimulated neurons (neurons cultured on CNT/Silk films with light applied, labeled CNT/silk light +) and two control groups, including neurons cultured on CNT/Silk films without application of light (labeled CNT/silk light -) and neurons cultured on silk films with application of light (labeled Silk light +). Representative traces are shown in **Figure 3d**. No significant fluorescence change was observed in the control groups. This result confirmed that the induced neuronal activity observed was due to the PA stimulation performed by the CNT/Silk films.

We next investigated the effect of laser pulse energy on PA stimulation. The duration of the pulse train in each stimulation was fixed at 5 ms. Three laser pulse energies, 23.5 µJ, 14.7 µJ and 8.8 µJ were applied to the CNT/Silk films to modulate neural activity. The PA stimulation successfully induced neural activity with laser pulse energies of 23.5 µJ and 14.7 µJ, but not with 8.8 µJ. Representative calcium trances are plotted in **Figure 3e**. Results from more neurons for each pulse energy are plotted as heatmaps in **Figure 3f-i**. For the laser pulse energy of 23.5 µJ and 14.7 µJ, 99.0% and 96.1% neurons in the illuminated area showed a fluorescence change above 10% after laser onset, respectively. But when the laser pulse energy was reduced to 8.8 µJ, the ratio was only 1.2%, which is more consistent with spontaneous action potential firing. The averages of maximum fluorescence change obtained from these three groups and from neurons cultured on the pure silk scaffolds as controls with pulse energy of 14.7 µJ are compared **in Figure 3j**. With the laser pulse energy of 23.5 µJ and 14.7 µJ, neurons showed an average maximum fluorescence change of 134% ± 56.9% and 87.3% ± 40.8%, significantly higher than 7.7% ± 16.9% from the silk only control group. When the pulse energy was reduced to 8.8 µJ, it was no longer able to evoke activation and the amplitude of the maximum fluorescence change was similar to the silk control group (n.s., $p = 0.982$). These results suggested that there was a threshold of laser pulse energy between 8.8 µJ and 14.7 µJ required for activation, given a pulse train duration of 5 ms. We had previously measured that the CNT/silk films generated PA waves with a peak pressure of 0.185 MPa for a laser pulse energy of 14.7 µJ. Therefore, we conclude that the threshold pressure for successful activation is close to 0.185 MPa, which is in the range of pressure (0.1-2 MPa) used for ultrasound neural stimulation [26]. Larger pulse energy led to PA waves with larger amplitudes (**Fig. 1e**) and induced larger calcium influxes, suggesting more action potential firings.

To further investigate whether CNT/silk scaffolds activate neurons reliably and repeatedly, we stimulated the same area 5 times sequentially, at 1 min intervals. Five repeated successful activations were observed for all 5 stimulations of the same neuron **(Figure 3k)**. We observed a decrease in max ΔF/F for each sequential stimulation, which could be attributed to calcium depletion[27] or spike frequency adaptation[28]. This result demonstrated the repeatability of PA stimulation of the CNT/silk scaffolds, and also confirmed that no damage was inflicted to the neurons by PA stimulation.

To confirm the minimal thermal effect associated with light conditions used for successful PA stimulation, we examined the surface temperature of the CNT/Silk scaffolds again for these specific conditions using a miniaturized ultrafast thermal probe **(Figure S5)**. The temperature increase was below 0.1°C, 0.15°C and 0.33°C for 8.8 μJ, 14.7 μJ and 23.5 μJ laser pulse energies, respectively. These temperature changes were much smaller than the previously reported threshold for thermally induced neural activation through infrared neural stimulation ($\Delta T > 5\ °C$)[29]. Therefore, neuronal activity stimulated by the CNT/Silk scaffolds was mainly due to the generated PA waves rather than a thermal effect. This minimal thermal effect means that CNT/Silk PA stimulation avoids potential thermotoxicity to neural tissue, compared with thermal-based neural stimulation.

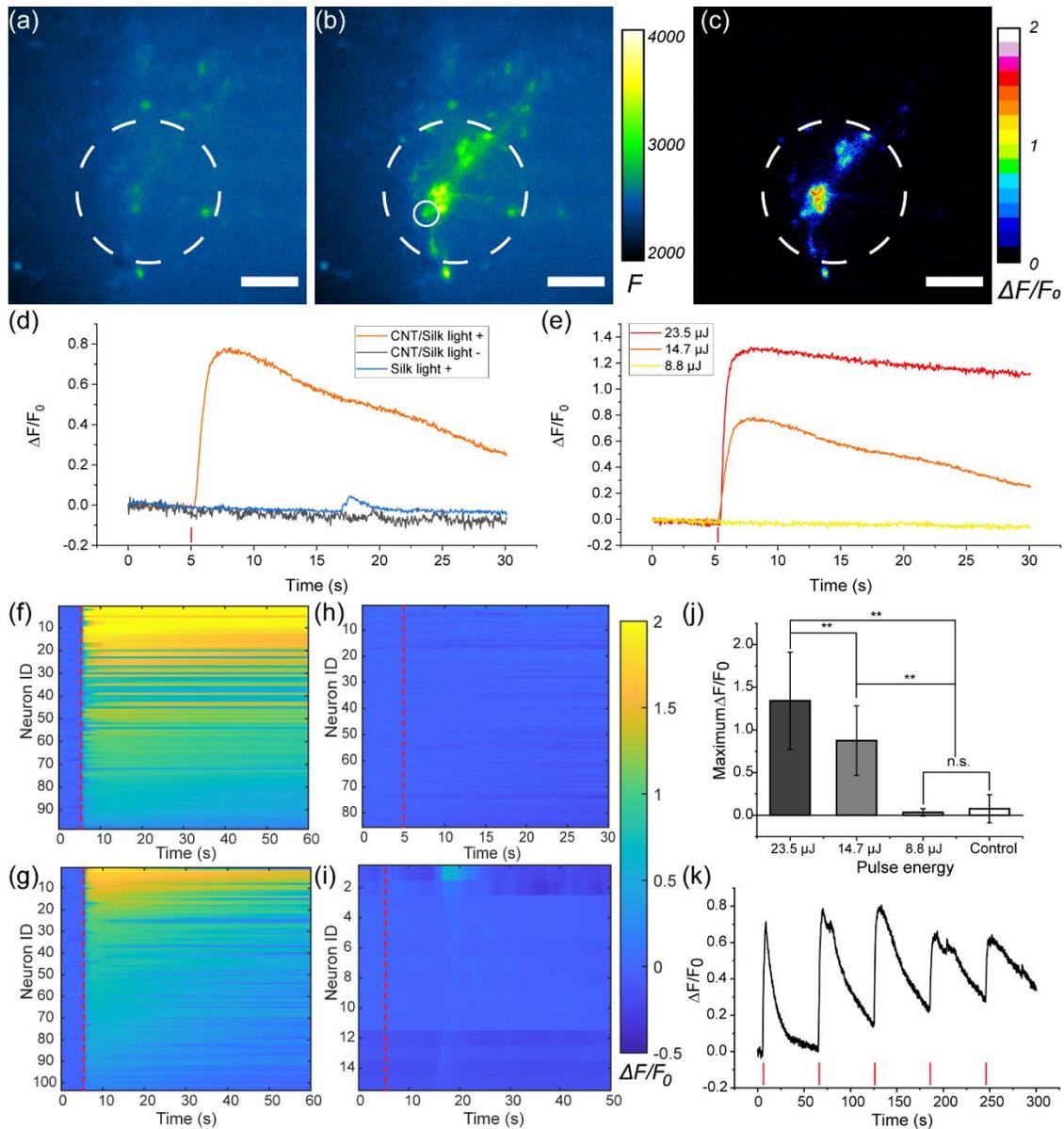

**Figure 3. PA waves generated by CNT/silk film stimulate neuronal activities in cultured primary neurons.** (a, b) Representative calcium images of GCaMP6f-transfected cortical neurons at DIV 14 cultured on CNT/silk film before (a) and after (b) the PA stimulation. Laser: 3 ns laser applied for a 5 ms duration with a 1.7 kHz repetition rate (corresponding to 8 pulses) and 14.7 µJ pulse energy. (c) Map of the maximum fluorescence change $\Delta F/F_0$ induced by the PA stimulation. Dashed lines: Illumination area. Scale bars: 100 µm. (d) Representative calcium traces of neuron cultured on CNT/Silk scaffold within the laser illumination area (CNT/silk light +, orange), out of the laser illumination area (CNT/silk light -, black) and cultured on silk scaffold within the laser illumination area (Silk light +, blue). (e) Representative calcium traces of neurons cultured on CNT/Silk film responding to PA stimulation with the pulse energy of 23.5 µJ (red), 14.7 µJ (orange)

and 8.8 µJ (yellow). (f-i) Colormaps of fluorescence change in neurons cultured on CNT/silk scaffold with laser pulse energy of 23.5 µJ (f), 14.7 µJ (g) and 8.8 µJ (h) and silk scaffold with laser pulse energy of 14.7 µJ (i). All PA stimulation was performed at 5 s and marked by the red dashed lines. The duration of each stimulation was fixed at 5 ms. (j) Average of maximum fluorescence intensity changes shown in (f-i). (k) Calcium trace of the neuron marked by the solid circle in (b) undergone repeated PA stimulation. Error bars represent standard deviation (n > 14, ** $p < 0.01$, n.s. $p = 0.982$, one-way ANOVA and Tukey's means comparison test).

**2.4 PA stimulations promote neurite extension**

After demonstrating the ability of CNT/Silk scaffolds for neural stimulation, we next evaluated the efficacy of PA stimulation on neurite extension. Whole DRG explants from E15 Sprague Dawley (SD) rat embryos were utilized as an ex vivo model. DRG explants are widely used to facilitate the study of nerve regeneration, because they allow for the evaluation of neurite outgrowth [30]. DRG explants with a diameter of approximately 300 µm were randomly selected and seeded onto either CNT/Silk scaffolds or a glass substrate as a control. Both substrate surfaces were coated with poly-D-lysine (50 µg/mL) and laminin (5 µg/mL) to improve DRG adhesion. DRGs in all groups were immersed in neurobasal medium supplemented with N2 and B27 but without neurotrophic factors. Using calcium imaging, we first confirmed that the PA waves generated by the CNT/Silk scaffold also stimulated DRG explants **(Figure. S6)**. All DRGs were grown for 2 days prior to stimulation. The stimulation conditions were the same as those used in cultured neurons (a laser train duration of 5 ms, 1.7 kHz repetition rate, a pulse energy of 14.7 µJ). The temperature increase associated with the stimulation was 0.15°C, indicating a negligible thermal effect on neurite growth [31].

To study the effect of stimulation on outgrowth, guided by previous electrostimulation and optogenetic stimulation work on neural growth, we applied PA stimulation for a total duration of 1 hour. During this 1-hour treatment, the laser pulse train of 5 ms every 2 minutes was applied to avoid potential desensitization and calcium depletion.

We designed the experiments to evaluate the effects of PA stimulation and substrates on neurite outgrowth. For this purpose, in addition to the PA-stimulated group (CNT/Silk light +), we tested three other control groups, including DRGs cultured on CNT/Silk without light illumination (CNT/Silk light -), DRGs cultured on glass with light illumination (glass light +) and DRGs cultured on glass without light stimulation (glass light -). All DRGs were grown for 2 days prior to stimulation, and fixed and stained at day 10 with Anti-Neurofilament 200 to label neurites and DAPI to label nuclei for immunofluorescence imaging and analysis (**Figure 4a-h**). To minimize human error during area counting, an algorithm for determining coverage area was developed **(Figure S7)**. High magnification confocal images demonstrated that the neurites and nuclei were stained by the corresponding antibodies. The neurite extensions of PA-stimulated DRGs were significantly promoted as compared to the DRGs in the control groups **(Figure 4i)**. For the PA-stimulated DRGs, the average growth area was $18.19 \pm 1.38$ mm$^2$, a 1.63-, 1.61- and 1.74-fold increase as compared to DRGs cultured on CNT/Silk without light illumination ($11.18 \pm 3.24$ mm$^2$), DRGs cultured on glass with light illumination ($11.31 \pm 1.37$ mm$^2$) and DRGs cultured on glass without light illumination ($10.48 \pm 5.06$ mm$^2$). These results confirmed that PA stimulation generated by the CNT/silk scaffolds had a strong positive effect on neurite outgrowth, and offers a new strategy to promote nerve regeneration.

To investigate the effect of PA stimulation dose on neurite outgrowth, we further examined the neurite extension of PA-stimulated DRGs varying the stimulation frequency from every 30 seconds to every 4 minutes, while maintaining the same stimulation conditions and the total duration of 1 hour (**Figure S8**). The average coverage area observed for groups stimulated every 30 s, every 1 min, every 2 min and every 4 min were 9.19 ± 2.95 mm$^2$, 13.15 ± 1.95 mm$^2$, 18.38 ± 1.38 mm$^2$ and 13.14 ± 4.14 mm$^2$, respectively. These results indicated that the positive effect of PA stimulation on neurite outgrowth reached a maximum when stimulation was applied every 2 minutes. For the group with the highest stimulation frequency at every 30 s, the average coverage area of DRGs (9.19 ± 2.95 mm$^2$) was slightly smaller than the control group (10.48 ± 5.06 mm$^2$).

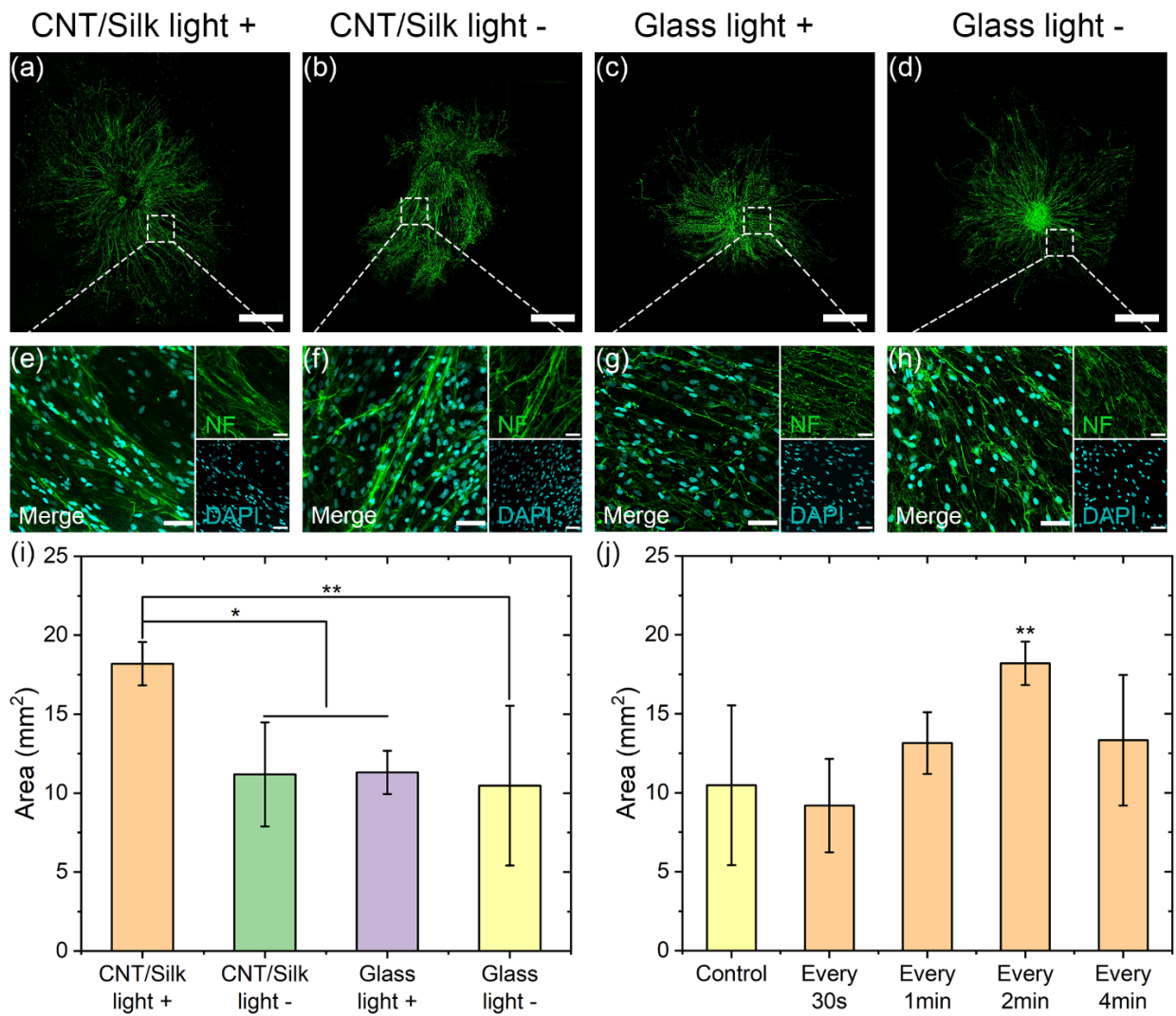

**Figure 4. PA stimulation promotes the neurite outgrowth.** (a-d) Representative confocal images of DRGs stained for neurofilament (green): DRG cultured on CNT/Silk film with laser illumination (CNT/Silk light +, a) and without laser illumination (CNT/Silk light -, b). DRG cultured on glass bottom dish with laser illumination (Glass light +, c) and without laser illumination (Glass light -, d). Scale bar: 1 mm. (e-h) High-resolution confocal images of DRGs stained with Anti-Neurofilament 200 (NF) for neurites (green) and DAPI for nucleus (cyan). Scale

bar: 50 μm. (i) The average neurite coverage area for DRGs in four groups. (j) The average neurite coverage area for PA stimulated DRGs with various stimulation frequency. Control: DRGs cultured on glass without light stimulation (glass light -). Laser train duration of 5 ms, a 1.7 kHz repetition rate, and the pulse energy of 14.7 μJ. All DRGs were allowed to grow for 2 days before stimulation, and were fixed at day 10. Error bars represent standard deviation (n = 5, ** $p < 0.01$, * $p < 0.05$, one-way ANOVA and Tukey's means comparison test).

To test whether PA stimulation increased the expression of neurotrophic factors, we measured the concentration of BDNF and nerve growth factor (NGF) 1 day after stimulation in PA-stimulated DRGs at the optimized stimulation frequency of every 2 min for 1 hour. DRGs in the other three control groups used above were also tested as controls **(Figure. 5)**. Both BDNF and NGF were measured by ELISA. PA-stimulated DRGs showed a concentration of BDNF of $86.52 \pm 17.07$ pg/ml, a 1.96-fold increase, compared to $44.20 \pm 22.31$ pg/ml measured for the control group where DRGs were cultured on glass without light stimulation. The two other control groups, DRGs cultured on CNT/Silk without light stimulation and DRGs cultured on glass with light stimulation, showed BDNF concentrations of $26.19 \pm 24.82$ pg/ml and $33.03 \pm 25.58$ pg/ml, close to the glass light- control. All four groups exhibited a similar expression level of NGF, and no significant difference was observed in PA-stimulated DRGs ($p > 0.05$).

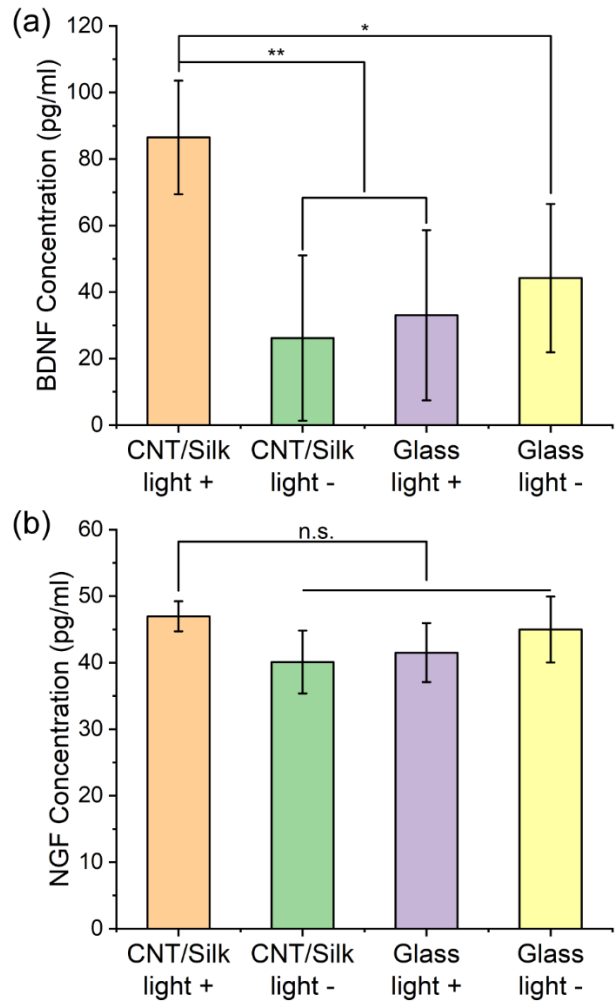

**Figure 5. Impact of PA stimulation on the expression of BDNF and NGF.** (a–b) The average concentrations of BDNF (a) and NGF (b) of PA-stimulated and unstimulated DRGs. Samples were collected 24 hours after PA stimulations. PA stimulations were provided every 2 minutes within the total duration of 1 hour. The stimulation condition was a laser train duration of 5 ms 1.7 kHz repetition rate, and the pulse energy of 14.7 µJ. (n = 5, ** $p < 0.01$, * $p < 0.05$, n.s. $p > 0.05$, one-way ANOVA and Tukey's means comparison test).

The increase in intracellular calcium caused by electrical stimulation and optogenetics has been shown to impact neural regeneration[5, 6b]. As a central regulator for neural growth, calcium serves multiple functions in the regeneration process[32] including regulating gene expression, reducing the membrane tension through activating proteases to cleave spectrin, or stimulating growth cone formation. These calcium-sensitive processes each have their own optimum calcium level, which could conceivably lead to an overall optimal level of calcium resulting in maximal neurite growth. Increasing calcium levels to reach the optimal level will increase growth, while increasing the level further above the optimal level will decrease the effect and could even lead to cell death[33]. With PA stimulation, the optimal frequency of every 2 min observed could be associated with the optimal calcium level for neurite outgrowth.

Impacting signaling pathways of neurotrophic factors is one mechanism through which stimulation promotes neural regeneration. Previous work demonstrated that increased neurite extension in response to electrical stimulation was correlated with changes in BDNF expression [34] and the externalization of its high affinity tropomysin-related kinase B (TrkB) receptor[35]. Besides the upregulation of BDNF, the concentration of NGF was observed to continuously increase both in electrically stimulated and unstimulated DRGs, but the elevation of NGF in stimulated DRGs was greater[5]. Our results indicated that PA stimulation with an optimized dose increased the expression of BDNF but did not significantly affect NGF. Here we suggest that PA stimulation promoted neurite outgrowth through two steps. First, PA stimulation elevated intracellular calcium concentration either by calcium influx following action potential firings, or directly opening the calcium channels. Then the elevated calcium concentration regulated a series of regeneration-related activities, including the expression of neurotrophic factors like BDNF.

2. Conclusions

We integrated PA capability into a silk-based neural scaffold using a nanocomposite approach, in which PEG-functionalized CNTs were dispersed in the silk matrix. Upon excitation by the 1030 nm pulsed laser, the CNT/Silk film generated a broadband PA wave, whose amplitude can be controlled by varying CNT concentration and laser energy. The pressure of the generated PA waves for successful neural stimulation is measured to be 0.185 MPa. Minimal thermal effects were associated with the illumination or the PA generation. The biocompatibility of this multifunctional scaffold was confirmed in vivo through H&E staining of histological slices of subcutaneously implanted scaffolds. The CNT/Silk scaffolds supported neural growth, and compared with control group, did not affect the morphology or viability. Through time-resolving calcium imaging, we demonstrate the reliable and repeatable neural stimulation function of the scaffold on the cultured cortical neurons with >94% neurons in the light illumination area activated. Using the DRG neurite extension model, CNT/Silk scaffolds are shown to promote neurite outgrowth of DRG explants by 1.74-fold and upregulate the expression of brain-derived neurotrophic factor (BDNF) compared with the control group.

Photoacoustic neural stimulation is an emerging method, and its mechanisms of action are not fully understood. For ultrasound neurostimulation using conventional transducers, several hypotheses have been proposed, including the activation of mechanosensitive ion channels[36], the transient mechanical disruption of the neural membrane including the opening of pores[37], and the induction of capacitive currents by intramembrane cavitation[38]. Our study indicates that transient disruption of the membrane and activation of ion channels are possibly involved in photoacoustic neural stimulation[14]. In the present work, we demonstrate that a PA material, specifically CNT/Silk, can be a new route to promote neural growth through neural stimulation. Compared to existing neural stimulation methods (electrical stimulation and optogenetics), this offers several advantages. First, it is a light-mediated method using NIR-II light. Because of its longer wavelength, NIR-II light has sufficient penetration depth in human tissue[39]. Thus, PA materials excited by NIR-II light potentially enable non-invasive postoperative stimulation. Second, compared with optogenetic stimulation, PA material-based neural stimulation does not require genetic transfection. As a foreign protein antigen, either the opsin itself or the expression method, like viral transfection, may raise technical and ethical issues for human applications[40]. Also, our work demonstrated that the photoacoustic capability can be integrated through a hydrogel nanocomposite approach. Owing to the design flexibility, this approach adds new light-responsiveness into hydrogel materials. Among current strategies, photothermal is a major mechanism that enables the light-responsive functions of hydrogel scaffolds in numerous applications[41]. However, in tissue engineering applications[42], damage due to heating surrounding tissues has to be considered. The PA-mediated light-responsiveness eliminates thermal toxicity because of the negligible temperature increase.

**Experimental Section/Methods**

*Silk Fibroin Preparation*: The silk fibroin solution used for the silk roll fabrication was prepared using published protocols[43]. Briefly, pieces of *Bombyx mori* cocoons were boiled in 0.02 M aqueous $Na_2CO_3$ for 30 min. The degummed silk was extensively rinsed in distilled water, dried overnight, and dissolved in 9.3 M LiBr at 60°C for 4 h. The silk/LiBr solution was dialyzed against distilled water for 2 d with ten changes of water. The solution was centrifuged at 9,000 rpm for 2 × 20 min. For the subsequent silk roll fabrication, the silk concentration was determined using an analytical balance by evaporating water from a solution of known weight and weighing the remaining solid.

*Preparation of DSPE-PEG 2000 Functionalized CNT*: Multiwall CNT (<8 nm OD, 2–5 nm ID, Length 0.5-2 µm, VWR Inc., NY, USA) aqueous solution (2 mg mL$^{-1}$) and DSPE-PEG 2000 (88120, Avanti Polar Lipids, AL, USA) aqueous solution (5 mg mL$^{-1}$) were mixed at a volume ratio of 1:2 and sonicated 3 hours in a bath sonicator at room temperature. Unfunctionalized CNT was first removed by centrifugation at 10k and 12k rpm for 10 minutes twice, respectively. Then the unbound surfactant was thoroughly removed by repeated filtration through 100 kDa filters (Vivaspin 6, Sartorius) at 3.5k rpm for 3 × 10 min. The concentration was determined by the absorbance at 500 nm of a diluted sample compared with the standard curve.

*Fabrication of CNT/Silk film and CNT/Silk Roll Scaffolds*: The CNT/Silk films were fabricated on a glass substrate by drop-casting 125 µL premixed silk (2%) and CNT (0-100 µg mL$^{-1}$) solution per cm$^2$ and then immersed in an 80% methanol solution to ensure insolubility. The CNT/Sill roll scaffold was fabricated by a bilayer structure following our previously reported method[21]. A bilayer film was first fabricated on a cover glass by drop-casting 41.67 µL agarose solution (1%) per cm$^2$ and 125 µL premixed silk (2%) and CNT (0-100 µg mL$^{-1}$) solution per cm$^2$ sequentially. After drying, the agarose/silk bilayer film was cut to desired size and aspect ratios. After immersion in 80% methanol solution, the self-folding CNT/Silk rolls were achieved by peeling the bilayer film off the glass substrate and immersing in water. All scaffolds were disinfected with 70% ethanol and UV light overnight.

*Characterization of CNT/Silk scaffolds*: For generating PA, a 1030 nm Q-switch laser with a pulse width of 3 ns was delivered to the scaffold though a multimodal optical fiber with 200 um core diameter (Thorlabs Inc.). An ultrasound transducer (10 MHz, XMS-310-B, Olympus, MA) with a preamplifier (0.2–40 MHz, 40 dB gain, Model 5678, Olympus, USA) was utilized to detect the acoustic waves. A digital oscilloscope (DSO6014A, Agilent Technologies, CA) was used to display and record the converted electrical signals from the transducer. The scaffold and ultrasound transducer were both immersed in deionized water to reduce the loss during the wave propagation and mimic the real application environment. The thermal effect generated by the scaffold was characterized by monitoring the change of surface temperature of scaffold in dry state. Thermal camera (A400, FLIR® Systems, Inc.) and thermal probe (DI-245, DATAQ Instruments, Inc., OH) were used to measure the temperature change.

*Cortical neuron and dorsal root ganglia (DRG) isolation and culture*: All experimental procedures complied with all relevant guidelines and ethical regulations for animal testing and research established and approved by Institutional Animal Care and Use Committee (IACUC) of Boston University (PROTO201800534). Both primary cortical neuron and DRGs were isolated from embryonic day 15 (E15) Sprague-Dawley rat embryos of either sex (Charles River Laboratories, MA).

To obtain cortical neurons, cortices were dissected and digested in TrypLE Express (Thermofisher scientific). Then the neurons were plated on poly-D-lysine (50 μg mL$^{-1}$, Thermofisher scientific) coated CNT/Silk scaffolds, silk scaffolds or glass. Neurons were first cultured with a seeding medium comprised of 90% Dulbecco's Modified Eagle Medium (Thermofisher scientific) and 10% fetal bovine serum (Thermofisher scientific) and 1% GlutaMAX (Thermofisher scientific), which was then replaced 24 h later by a growth medium comprised of Neurobasal Media supplemented with 1 × B27 (Thermofisher scientific), 1 × N2 (Thermofisher scientific) and 1 × GlutaMAX (Thermofisher scientific). AAV9.Syn.Flex.GCaMP6f.WPRE.SV40 virus (Addgene, MA) was added to the cultures at a final concentration of 1 μL mL$^{-1}$ at day 5 in culture for GCaMP6f expression. Half of the medium was replaced with fresh growth medium every 3-4 days. Cells cultured in vitro for 10-14 days were used for PA stimulation experiments.

To extract DRGs, the spinal cord was exposed and individual DRG explants were trimmed of nerve roots and additional connective tissue and placed in a container of growth medium comprised of Neurobasal Media supplemented with 1 × B27 (Thermofisher scientific), 1 × N2 (Thermofisher scientific), 1 × GlutaMAX (Thermofisher scientific) and 50 U/mL penicillin/streptomycin (Thermofisher scientific). Scaffold or glass were coated with poly-D-lysine (50 μg mL$^{-1}$, Thermofisher scientific) and laminin (5 μg mL$^{-1}$, Corning) and rinsed with growth medium prior to seeding. DRGs were directly seeded on the scaffold or glass substrate. They were allowed to attach the substrate for 4 hours with 20 μL growth medium for each DRG explant. After 4 hours, DRGs were covered by the growth medium. Half of the medium was replaced with fresh growth medium every 3 days. All DRGs were grown for 48 hours prior to applying PA stimulation and fixed 10 days following seeding.

*MTS viability assay*: Neuronal cell viability was determined using the 3-(4,5-dimethylthi-azol-2-yl)-5-(3-carboxymethoxyphenyl)-2-(4-sulfophenyl)-2H-tetrazolium reduction assay (MTS assay, Abcam) following the manufacturer's instructions. Unlike the MTT assay, the colored formazan dye generated in the MTS assay was soluble in cell culture medium. To avoid interference from the color of CNT/Silk scaffolds, the final solution was transferred into a new 96-well plate and the absorbance was measured at a wavelength of 490 nm using a microplate reader (SpectraMax i3x, Molecular Devices).

*Histological analysis*: All animal experiments were performed under the guidance and approval of the Institutional Animal Care and Use Committee (IACUC) of Tufts University (protocol number: M2019-121). Ten female FVB/NJ mice aged 10 weeks were randomly separated into 2 groups. Group 1 was euthanized 3 days after implantation, to assess the acute immune response. Group 2 was euthanized 30 days after implantation to assess chronic inflammation. Two small longitudinal incisions (~ 0.5-1 cm) were made through the skin of each mouse. Sterile CNT/Silk roll scaffold were inserted in the resulting subcutaneous pouches. The incision was closed with surgical clips. Animals were monitored daily for the first three days after implantation, then weekly until euthanasia.

*In vitro neurostimulation:* Cortical neurons/DRGs were cultured on CNT/Silk scaffold. A Q-switched 1030-nm nanosecond laser with repetition rate of 1.7 kHz and the pulse width of 3 ns (Bright Solution, Inc. Calgary Alberta, CA) was delivered through a 200 μm core diameter multimodal optical fiber. The fiber position was controlled by a 3-D micromanipulator (Thorlabs, Inc., NJ, USA) to target an area of the scaffold. During the PA stimulation, the fiber was placed 50 μm approximately above the scaffold and the illumination area was calculated to be 0.05 $mm^2$. Calcium fluorescence imaging was performed on a lab-built wide-field fluorescence microscope based on an Olympus IX71 microscope frame with a 20x air objective (UPLSAPO20X, 0.75NA, Olympus, MA), illuminated by a 470 nm LED (M470L2, Thorlabs, Inc., NJ) and a dichroic mirror (DMLP505R, Thorlabs, Inc., NJ). Image sequences were acquired with a scientific CMOS camera (Zyla 5.5, Andor) at 20 frames per second. The fluorescence intensities, data analysis and exponential curve fitting were analyzed using ImageJ (Fiji) and Origin 2018.

*Immunocytochemistry*: Cortical neurons were fixed at day 7 with 4% (wt. %) paraformaldehyde (Thermofisher scientific) solution in phosphate buffered saline (PBS, Thermofisher scientific) for 20 min and permeabilized with 0.1% Triton X-100 (Sigma) in PBS for 10 min at room temperature. After 1 hour blocking with 5% Bovine Serum Albumin (BSA) in PBS at room temperature, cells were incubated with the primary antibody, 1:200 anti-Tau antibody (TAU-5) for 2 hours. Then cells were incubated with the secondary antibody, 1:500 Alexa Fluor 488 Goat anti-Mouse IgG (ab150113, Abcam) for 1 hour. Lastly, samples were incubated with 1:10,000 DAPI (D9542, Sigma) solution in PBS for 15 min.

    DRGs were fixed at day 10 with 4% (Wt. %) paraformaldehyde (Thermofisher scientific) solution in phosphate buffered saline (Thermofisher scientific) for 20 min and permeabilized with 0.1% Triton X-100 (Sigma) in PBS for 10 min at room temperature. After blocking overnight with 2.5% goat serum (Abcam) in PBS at 4°C, DRG samples were incubated with the primary antibody, 1:500 rabbit anti-neurofilament antibody (N4142, Sigma) solution in 2.5% goat serum solution in PBS for 1 hour. Then samples were incubated with the secondary antibody, 1:1000 Alexa Fluor 633 Goat anti-Rabbit IgG (A-21070, Thermofisher scientific) in 2.5% goat serum in PBS for 1

hour. Lastly, samples were incubated with 1:10,000 DAPI (D9542, Sigma) solution in PBS for 15 min.

*Confocal imaging and image analysis*: Confocal images were acquired by a laser scanning confocal microscope (FV3000, Olympus). The overview images were taken with an air-immersion 10X objective (Olympus) and stitched by ImageJ (Fiji). The detail high magnification images were taken with 20X objective (Olympus). The coverage area of DRGs were calculated by an analysis algorithm written in MATLAB (Mathworks, Natick, MA). The stitched confocal images were firstly converted into a binary image and the boundary of DRGs were marked along 360 radial lines separated by 1º. A polygon was then constructed based on these boundary points. Finally, the coverage area was derived from the area of the polygon and scale conversion.

*ELISA assay*: Cell extractions of PA stimulated DRGs and DRGs in all control groups were collected 24 hours following stimulation by a cell lysis buffer (FNN0011, Thermofisher scientific). The concentrations of brain-derived neurotrophic factor (BDNF) and nerve growth factor (NGF) were measured by the ELISA assay kits (ERBDNF and ERNGF, Thermofisher scientific) following instructions provided. Briefly, experiment samples and standard samples were incubated with pre-coated 96-well plate for 2.5 hours at room temperature. Then biotin conjugate, Streptavidin-HRP, TMB substrate and stop solution were added and incubated successively. Optical absorbance was measured using a microplate reader (SpectraMax i3x, Molecular Devices) at 450 nm, and the concentration was calculated according to the standard curve.

*Statistical analysis*: Data shown are mean ± standard deviation (SD). For the comparison on neurons per area, total neurite length per area and cell viability, N = 4 samples were analyzed using a one-way ANOVA with Tukey's post-hoc test. For the comparison on DRG coverage area and the concentration of neurotrophic factors, N = 5 samples were analyzed using a one-way ANOVA with Tukey's post-hoc test. P values were determined as: n.s.: non-significant, $p > 0.05$; *: $p < 0.05$; **: $p < 0.01$. Fluorescence images were analyzed using ImageJ (Fiji). Biocompatibility, maximum fluorescence change, DRG coverage area and concentrations of neurotrophic factors were analyzed and plotted using Origin. The coverage area of DRGs were calculated by a custom algorithm in MATLAB.

**Acknowledgements**
We thank Boston University Clinical and Translational Science Institute Pilot Award (funded by NIH 1UL1TR001430) and the NIH (P41EB027062) for support of this work.

**Table of Contents**

*Nan Zheng, Vincent Fitzpatrick, Ran Cheng, Linli Shi, David L. Kaplan, and Chen Yang*[*]

N. Zheng, Prof. C. Yang

Division of Materials Science & Engineering, Boston University, Boston, MA 02215, USA

E-mail: cheyang@bu.edu

Dr. V. Fitzpatrick, Prof. D. L. Kaplan

Department of Biomedical Engineering, Tufts University, Medford, MA 02215, USA

E-mail: david.kaplan@tufts.edu

R. Cheng, L. Shi

Department of Chemistry, Boston University, Boston, MA 02215, USA

Multifunctional CNT/silk neural scaffolds are achieved by integrating photoacoustic function through these nanocomposite-hydrogels. Using infrared nanosecond pulsed laser, these biocompatible scaffolds enable reliable and repeatable photoacoustic neural stimulation functions with in vivo biocompatability. Photoacoustic stimulation and these new scaffolds promote neurite outgrowth, providing a new route for the repair of nerve injury through stimulating neuronal activities.

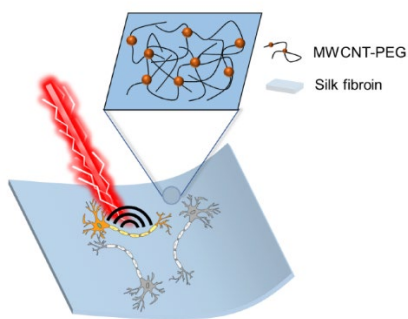

**Supplementary materials**

*Nan Zheng, Vincent Fitzpatrick, Ran Cheng, Linli Shi, David L. Kaplan[*], and Chen Yang[*]*

N. Zheng, Prof. C. Yang

Division of Materials Science & Engineering, Boston University, Boston, MA 02215, USA

E-mail: cheyang@bu.edu

Dr. V. Fitzpatrick, Prof. D. L. Kaplan

Department of Biomedical Engineering, Tufts University, Medford, MA 02215, USA

E-mail: david.kaplan@tufts.edu

R. Cheng, L. Shi

Department of Chemistry, Boston University, Boston, MA 02215, USA

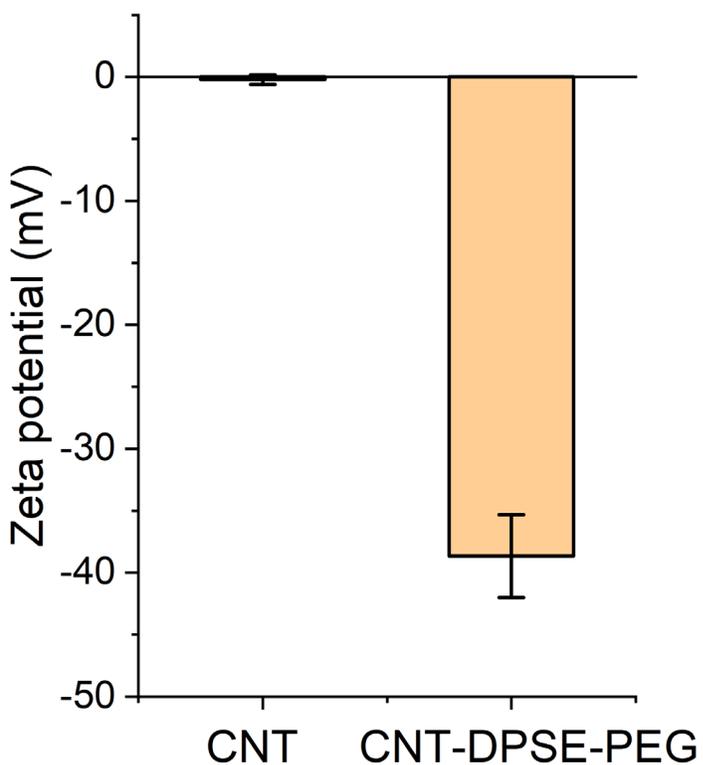

**Figure S1.** The zeta-potential of raw MWCNT and DPSE-PEG functionalized MWCNT solution. Negative charges were introduced after functionalization.

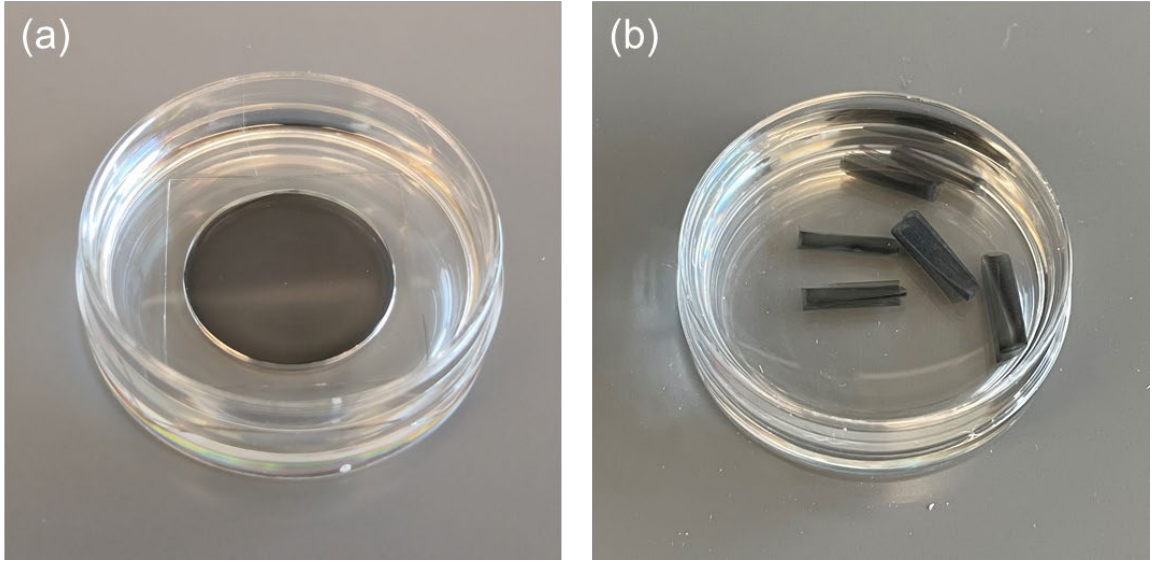

**Figure S2.** The CNT/silk scaffold was fabricated through a cast-and-dry method into a film with thickness of 32.54 ± 1.98 μm (a) or a 3D roll structure (b) through the self-folding strategy.

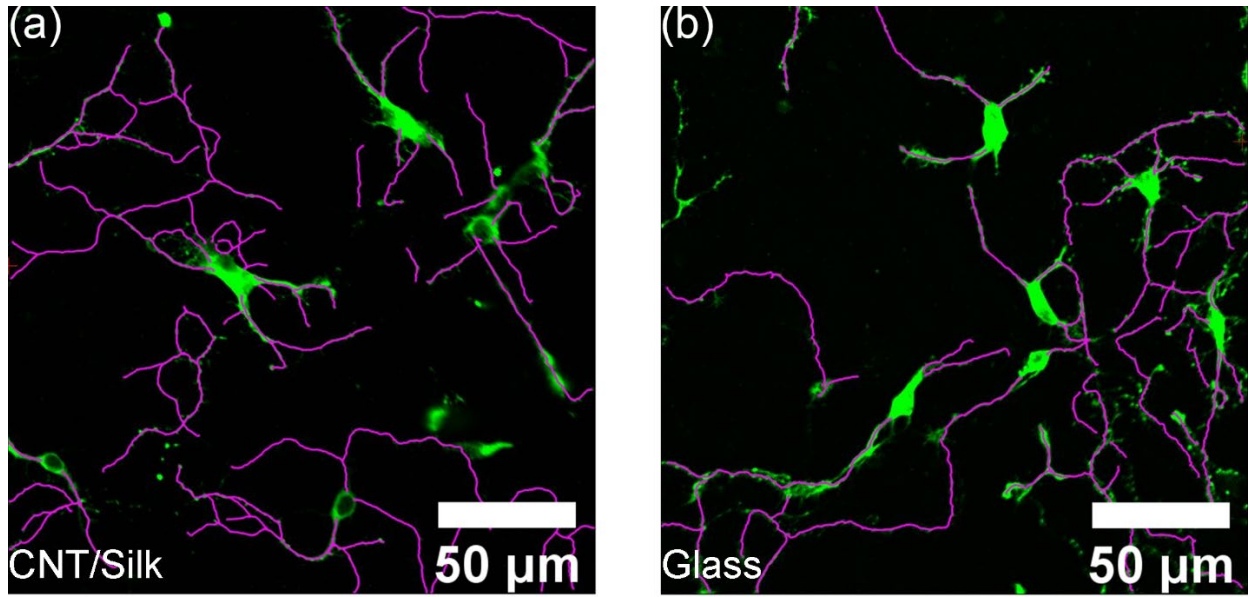

**Figure S3.** Neurite tracing of immunofluorescence images of cortical neurons cultured on CNT/silk film(a) and glass substrate(b). Neurons were labeled with anti-Tau antibody (green).

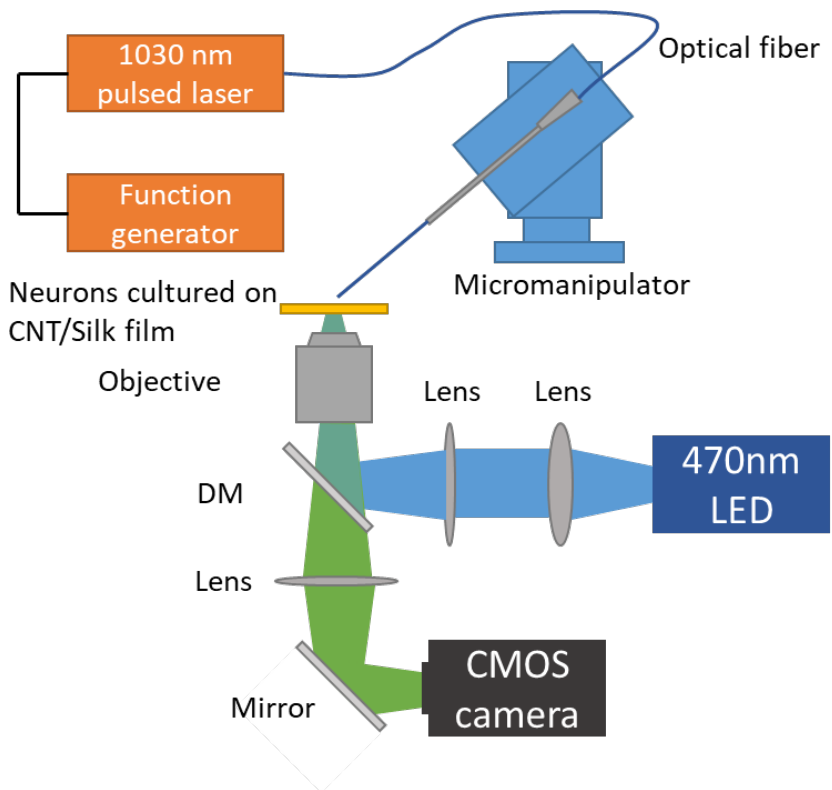

**Figure S4.** Calcium imaging was performed on an inverted wide field fluorescence microscope to monitor neuronal activities. Pulsed laser light was delivered through a multimode optical fiber and the illumination area was controlled by a micromanipulator. DM: dichroic mirror.

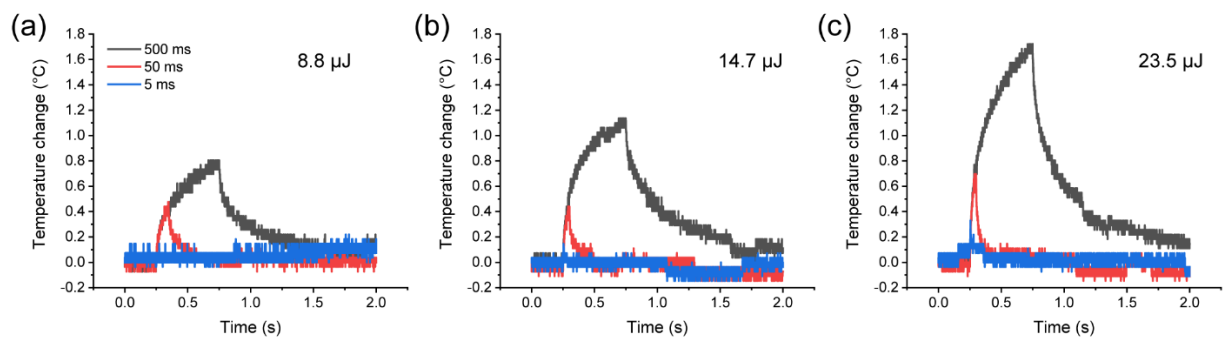

**Figure S5.** Heat profile of CNT/Silk scaffold under the laser illumination. Three pulse energy, 8.8 µJ (a), 14.7 µJ(b) and 23.5 µJ(c) were examined with duration of 5 ms, 50 ms and 500 ms.

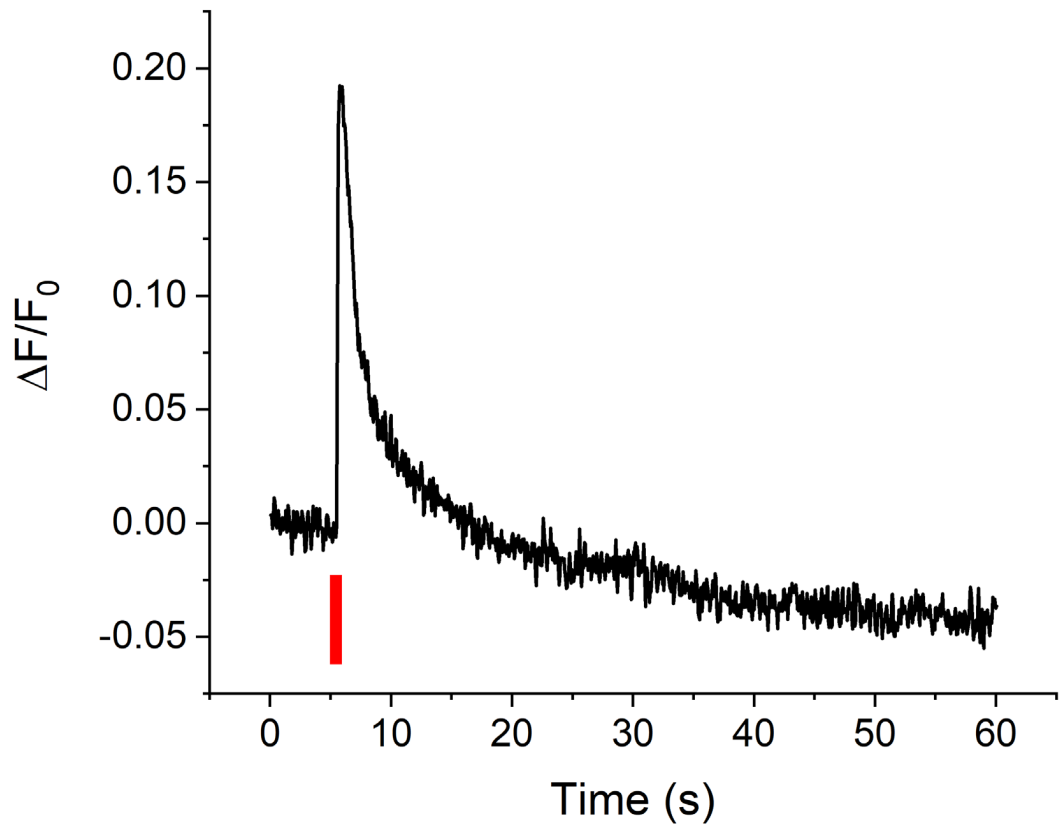

**Figure S6.** Calcium trace of a GCaMP6f transfected DRG at DIV 10 under the PA stimulation with the pulse energy of 14.7 µJ and the duration of 5 ms. The stimulation was triggered at 5 s.

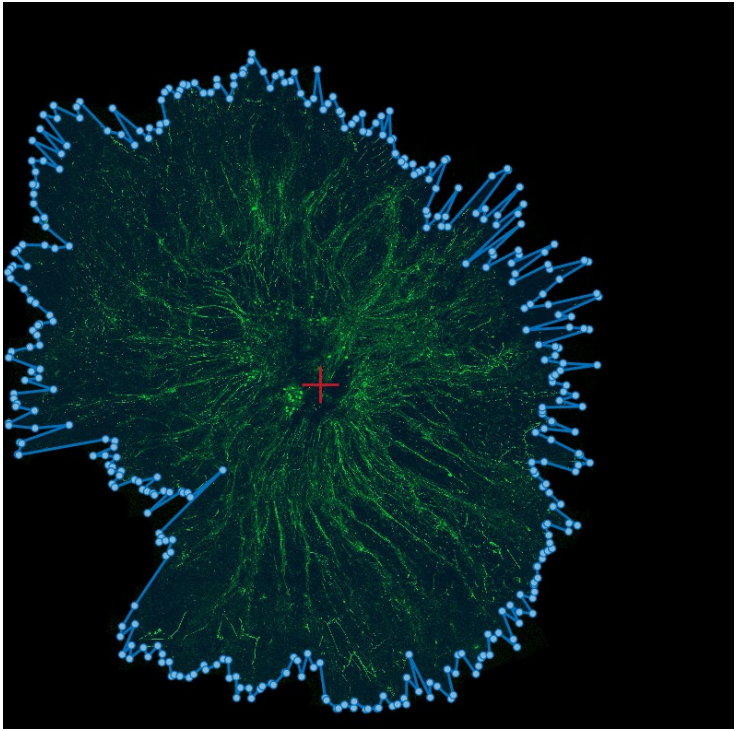

**Figure S7.** The coverage area of DRGs was determined by a computer algorithm to minimize the artificial error. The polygon was constructed by 360 endpoints along each direction separated by 1º.

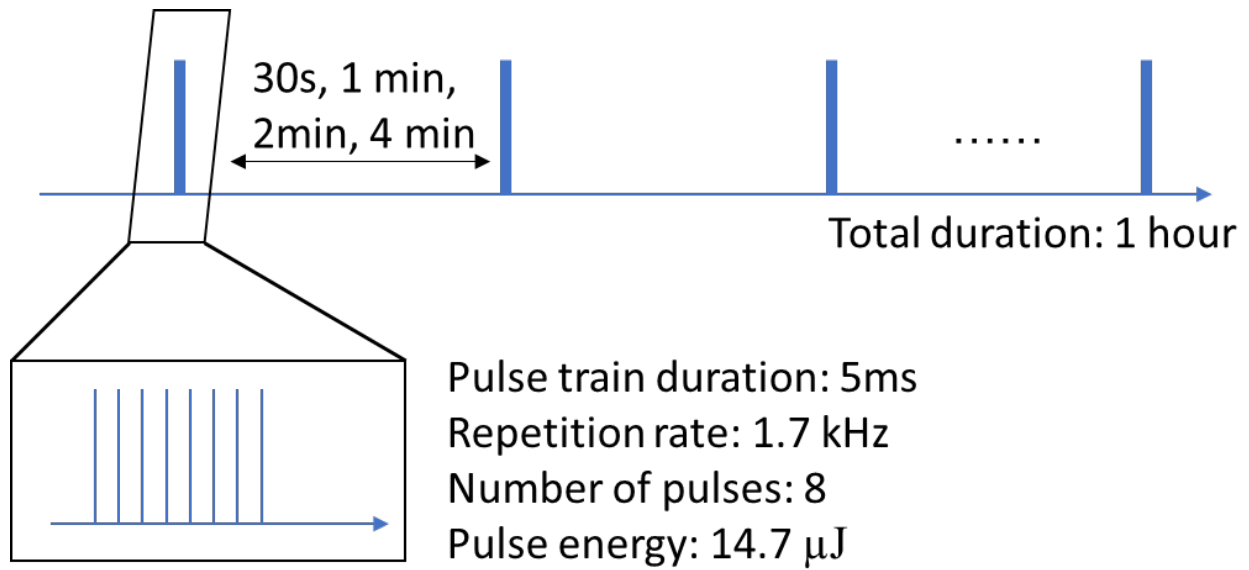

**Figure S8.** The schematic of laser pulse train used for investigating the effect of PA dosage. The total duration was fixed at 1 hour. The stimulation frequency was varied from every 30 seconds to every 4 minutes. Within each 5 ms pulse train, 8 pulses with pulse energy of 14.7 µJ were provided.